\documentclass{amsart}
\usepackage{amsfonts,amsmath,amssymb}

\newtheorem{theorem}{Theorem}
\theoremstyle{plain}

\newtheorem{definition}{Definition}
\newtheorem{example}{Example}

\numberwithin{equation}{section}

\begin{document}
\title[]{Restructuring of Discreate Logarithm Problem and ElGamal
Cryptosystem by Using the Power Fibonacci Sequence Module $M$}
\author{Cagla OZYILMAZ}
\address{Ondokuz May\i s University, Samsun, Turkey}
\email{cagla.ozyilmaz@omu.edu.tr}
\urladdr{}
\thanks{}
\author{Ayse NALLI}
\curraddr{Karab\"{u}k University, Karab\"{u}k, Turkey}
\email{aysenalli@karabuk.edu.tr}
\urladdr{}
\thanks{}
\author{}
\address{\"{y} }
\urladdr{}
\date{}
\subjclass{}
\keywords{Asymmetric Cryptography, Discrete Logarithm Problem, Power
Fibonacci Sequences, ElGamal Cryptosystem}
\dedicatory{}
\thanks{}

\begin{abstract}
In this paper, we have studied on adapting to asymmetric cryptography power
Fibonacci sequence module $m$ . To do this, We have restructed Discreate
Logarithm Problem which is one of mathematical difficult problems by using
power Fibonacci sequence module $m$ and by means of this sequences, we have
made the mathematical difficult problem which is used only in prime modules
is also useful for composite modules. Then we have constructed cryptographic
system based on this more difficult problem which we have rearranged. Hence,
we have obtained a new cryptosystem as ElGamal Cryptosystem. Lastly, we have
compared that ElGamal Cryptosystem and a new cryptosystem which we
constitute in terms of cryptography and we have obtained that a new
cryptosystem is more advantageuos than ElGamal Cryptosystem.
\end{abstract}

\maketitle

\section{Introduction}

The fundamental objective of cryptography is to enable two people, usually
referred to as Alice and Bob, to communicate over an insecure channel in
such a way that an opponent, Oscar, can't understand what is being said.
This channel could be a telephone line or computer network, for example. The
information that Alice wants to send to Bob, which we call ` plaintext ',
can be English text, numerical data, or anything at all- its structure is
completely arbitrary. Alice encrypts the plaintext, using a predetermined
key, and sends the resulting ciphertext over the channel. Oscar, upon seeing
the ciphertext in the channel by eavesdropping, can't determine what the
plaintext was; but Bob, who knows the encryption key, can decrypt the
ciphertext and reconstruct the plaintext.

\bigskip \bigskip These ideas are described formally using the following
mathematical notation.

\begin{definition}
A crptosystem is a five -- tuple$~(P,C,K,E,D)$ where the following
conditions are satisfied:

1. $P$ is a finite set of possible plaintexts;

2. $C$ is a finite set of possible ciphertexts;

3. $K$ is a finite set of possible keys;

4. For each $K\in K$, there is an encryption rule $e_{K}\in E~$and a
corresponding decryption rule $d_{K}\in D$. Each $e_{K~}:P~\longrightarrow
~C~$and ~$d_{K~}:C~\longrightarrow ~P~$are functions such that ~$%
d_{K~}(e_{K~}~(x))~=~x~~$for every plaintext element$~x\in ~P$\cite{stinson}.
\end{definition}

Fundamentally, there are two types of cryptosystems based on the manner in
which encryption-decryption is carried out in the system

\textbullet \qquad Symmetric Cryptography (Secret key cryptosystems)

\textbullet \qquad Asymmetric Cryptography (Public key cryptosystems)

The main difference between these cryptosystems is the relationship between
the encryption and the decryption key. Logically, in any cryptosystem, both
the keys are closely associated. It is practically impossible to decrypt the
ciphertext with the key that is unrelated to the encryption key. Algorithms
for symmetric cryptography, such as DES \cite{National}, use a single key
for both encryption and decryption and algorithms for asymmetric
cryptography, such as the RSA \cite{Rivest} and ElGamal cryptosystem\cite%
{ElGamal}, use different keys for encryption and decryption.

At the very heart of cryptography is the notion of one way function, which
was shown to be necessary and sufficient for many cryptographic primitives.

A one-way function(OWF) is a function $f\ \ $such that for each$~x$ in the
domain of \ $f$, it is easy to compute $f(x)$; but for essentially all $y~$%
in the range of ~$f$, it is computationally infeasible to find any~$x~$ such
that $y~=$ $f(x)$.

The following are two examples of candidate one-way functions.

1. OWF multiplication of large primes: For primes $p~$and $q$, $%
f(p,~q)~=~pq~ $is a one-way function: given $p~$and~$q$ ,computing $n~=~pq~$%
is easy; but given $n$, finding $p~$and~$q$ is difficult. The difficult
direction is known as the integer factorization problem, RSA and many other
cryptographic systems rely on this example.

2. OWF exponentiation in prime fields: Given a generator $\alpha ~$of ~$%
\mathbb{Z}
_{p}^{\ast }~$for most appropriately large prime $p$, $f(a)~=~\alpha
^{^{a}~}({mod}~p)~$is a one-way function. $f(a)~$is easily computed
given $\alpha $, $a$, $p$ and ; but for most choices $p~$it is difficult,
given ( $y$; $\alpha $; $p$), to find~$a$ an in the range $1\leq ~a\leq p-1~$%
such that $\alpha ^{^{a}~}({mod}~p)=~y$. The difficult direction is
known as the Discrete Logarithm problem.

However, a one-way function is not sufficient for public-key cryptography if
it is equally hard for the legitimate receiver and the adversary to invert.
So rather, we need a trapdoor one-way function. A trapdoor one-way function
is a one-way function where the inverse direction is easy, given a certain
piece of information (the trapdoor), but difficult otherwise. Public-key
cryptosystems are based on trapdoor one-way functions. The public key gives
information about the particular instance of the function; the private key
gives information about the trapdoor \cite{Zhu}.

\qquad Now, we cite public-key cryptosystems based on the Discrete Logarithm
problem. The first and best-known of these is the ElGamal Cryptosystem.
ElGamal proposed a public-key cryptosystem which is based on the Discrete
Logarithm problem in ( $%
\mathbb{Z}
_{p}^{\ast }$, $.$). The encryption operation in the ElGamal Cryptosystem is
randomized, since ciphertext depends on both the plaintext $x~$and on the
random value $k$ chosen by Alice. Hence, there will be many ciphertexts that
are encryptions of the same plaintext.

ElGamal Cryptosystem is presented below:

Let $p~$be a prime number such that the Discrete Logarithm problem in ( $%
\mathbb{Z}
_{p}^{\ast }$, $.$) is infeasible, and let $\alpha \in $ $%
\mathbb{Z}
_{p}^{\ast }~$be a primitive element. Let $P~=~%
\mathbb{Z}
_{p}^{\ast }$, $C~=~%
\mathbb{Z}
_{p}^{\ast }\times 
\mathbb{Z}
_{p}^{\ast }~$, and define $K~=~\{\left( p,~\alpha ,~a,~\beta \right)
:~\beta ~\equiv \alpha ^{^{a}~}({mod}~p)\}~$. The values $p,~\alpha
,~\beta ~$are the public key, and $a$ is the private key. For $K~=\left(
p,~\alpha ,~a,~\beta \right) $, and for a (secret) random number $k\in 
\mathbb{Z}
_{p-1}$, define $e_{K~}~(x,k)~=~\left( y_{1}~,y_{2}\right) ~$, where

\begin{equation*}
y_{1} =\alpha ^{^{k}~}({mod}~p) \\
y_{2} = x\beta ^{^{k}~}({mod}~p)
\end{equation*}

For $y_{1}~,y_{2}\in 
\mathbb{Z}
_{p}^{\ast }$, define $d_{K~}~(y_{1}~,y_{2})~=y_{2}\left( y_{1}~^{a}\right)
^{-1}({mod}~p).$

A small example will illustrate the computations performed in the ElGamal
Cryptosystem.

Suppose $p=2579~$and $\alpha =~2$. 

$\alpha ~$is a primitive element module $p$ . Let $a=~765$, so$~\beta
~=2^{765}({mod}~2579)=949.$

Now, suppose $~$that Alice wishes to send the message $x=1299$ to Bob. Say $%
k=853$ is the random integer she chooses. Then she computes

\begin{multline*}
y_{1} =\alpha ^{^{k}~}({mod}~p)=2^{^{853}~}({mod}~2579)=435 \\
y_{2} =x\beta ^{^{k}~}({mod}~p)=1299.949^{853~}({mod}~2579)=2396
\end{multline*}

Alice sends $y=\left( y_{1}~,y_{2}\right) =\left( 435~,2396\right) ~$to Bob.

When Bob receives the ciphertext $y=\left( 435~,2396\right) $, he computes

\begin{equation*}
x=y_{2}\left( y_{1}~^{a}\right) ^{-1}({mod}~p)=2396.\left(
435^{765}\right) ^{-1}({mod}~2579)=1299
\end{equation*}

which was the plaintext that Alice encrypted\cite{stinson}.

\section{Power Fibonacci Sequences}

\bigskip Let $G~$be a bi-infinite integer sequence satisfying the recurrence
relation $G_{n}~=G_{n-1}+G_{n-2}$. If$~\ G\equiv 1,\alpha ,\alpha
^{2},\alpha ^{3},...({mod}~m)~$for some modulus~$m$ , then $G~$is
called a power Fibonacci sequence modulo \cite{Ide}.

\begin{example}
Modulo $m=19$, there are two power Fibonacci sequences: 1, 15, 16, 12, 9, 2,
11, 13, 5, 18, 4, 3, 7, 10, 17, 8, 6, 14, 1, 15. . . and 1, 5, 6, 11, 17, 9,
7, 16, 4, 1, 5, . . .
\end{example}

Curiously, the second is a subsequence of the first. Ide and Renault, in
their paper, obtained for modulo $5$ there is only one such sequence (1, 3,
4, 2, 1, 3, ...), for modulo $10$ there are no such sequences, and for
modulo $209$ there are four of these sequences. Thus, they obtained the
following theorem.

\begin{theorem}
There is exactly one power Fibonacci sequence modulo $5$. For $m\neq 5$,
there exist power Fibonacci sequences modulo $m~$precisely when $m~$has
prime factorization $m~=~p_{1}^{e_{1}}.p_{2}^{e_{2}}...p_{k}^{e_{k}}~$or $%
m~=~5p_{1}^{e_{1}}.p_{2}^{e_{2}}...p_{k}^{e_{k}}$, where each$~p_{i}$ $%
\equiv \pm 1({mod}~10)$; in either case there are exactly $2^{k}~$power
Fibonacci sequences modulo $m$\cite{Ide}.
\end{theorem}

In this paper, we have examined power Fibonaci sequences modulo $m~$and we
have obtained that a power Fibonaci sequence modulo $m~$constitutes a cyclic
and multiplicative group whose order $~$is a divisor of $\varphi (m)~$at the
same time. That is, If $G\equiv 1,\alpha ,\alpha ^{2},\alpha ^{3},...({%
mod}~m)$ is a power Fibonacci sequence module $m$, the power Fibonacci
sequence constitutes a cyclic and multiplicative group whose generator is $%
\alpha $ . This sequence is a subgroup of $~%
\mathbb{Z}
_{m}^{\ast }$( $~%
\mathbb{Z}
_{m}^{\ast }=~\{x\in 
\mathbb{Z}
_{m}:(x,m)=1~\}$). In addition when we examine to mathematical difficult
problems which are used asymmetric cryptography, we get a generator is
necessary for the Discrete Logarithm problem. When we have thought all these
things together, we have obtained that we can rearrenge Discrete Logarithm
problem by using the power Fibonacci sequence module $m~$and so we can
rearrenge ElGamal Cryptosystem based on the Discrete Logarithm problem.
Thus, we have made that the Discrete Logarithm problem and ElGamal
Cryptosystem which is used only in prime modulus is also usable for
composite modulus.

Moreover, there are two limits in the ElGamal cryptosystem. One is that the
plaintext must be less than $p-1$\cite{Hwang}. So then if $m~$is chosen a
composite number by using the power Fibonacci sequence module $m$, how does
this limit in the ElGamal cryptosystem change? To obtain the answer of this
question, firstly we define new Discrete Logarithm problem by using the
power Fibonacci sequence module $m~$and then we constitute a new ElGamal
Cryptosystem based on new Discrete Logarithm problem.

\begin{definition}
Given a generator $\alpha ~$of a chosen subgroup of~$~%
\mathbb{Z}
_{m}^{\ast }~$for most appropriately large $m~$( according to Theorem 1, we
have found most appropriately large module~$m$ ), $f(\lambda )~=~\alpha
^{^{\lambda }~}({mod}~m)~$is a one-way function. $f(\lambda )$ is
easily computed given $\lambda $,$\alpha $ , and $m$; but for most choices $%
m $ it is difficult, given ( $y$; $m$ ; $\alpha $), to find$~$an $\lambda ~$%
such that $\alpha ^{^{\lambda }~}({mod}~m)=~y$. We have called
difficult direction as the Discrete Logarithm problem with Power Fibonacci
Sequence module $m$.
\end{definition}

Now, we will obtain public-key cryptosystem based on the Discrete Logarithm
problem with Power Fibonacci Sequence module $m$. We have called the
cryptosystem as the ElGamal Cryptpsystem with Power Fibonacci Sequence
module $m$ .

\qquad

ElGamal Cryptpsystem with Power Fibonacci Sequence module $m$ is presented
below:

Let $m~$be a positive integer such that the Discrete Logarithm problem with
Power Fibonacci Sequence module $m~$in a chosen subgroup of ( $%
\mathbb{Z}
_{m}^{\ast }$, $.$) ( $m~$is one of modulus for which power Fibonacci
sequences exist) is infeasible, and let $\alpha \in ~\left(
the~chosen~subgroup~of~%
\mathbb{Z}
_{m}^{\ast }\right) ~$be a primitive element(generator). Let

$P~=~%
\mathbb{Z}
_{m}/~\{0\}$, $C~=~\left( the~chosen~subgroup~of~%
\mathbb{Z}
_{m}^{\ast }\right) \times 
\mathbb{Z}
_{m}/~\{0\}~$, and define $K~=~\{\left( m,~\alpha ,~\lambda ,~\beta \right)
:~\beta ~\equiv \alpha ^{^{\lambda }~}({mod}~m)\}~$. The values$%
~m,~\alpha ,~\beta ~$are the public key, and $\lambda $ is the private key.
For $K~=\left( m,~\alpha ,~\lambda ,~\beta \right) $, and for a (secret)
random number $k\in 
\mathbb{Z}
_{the~order~of~the~chosen~subgroup~}$, define $e_{K~}~(x,k)~=~\left(
y_{1}~,y_{2}\right) ~$, where

\begin{equation*}
y_{1} =\alpha ^{^{k}~}({mod}~m) \\
y_{2} =x\beta ^{^{k}~}({mod}~m)
\end{equation*}

For $\left( y_{1}~,y_{2}\right) \in ~C$, define $%
d_{K~}~(y_{1}~,y_{2})~=y_{2}\left( y_{1}~^{\lambda }\right) ^{-1}({mod}%
~m).$

Thus, if we look closely, while $\alpha \in 
\mathbb{Z}
_{p}^{\ast }$, the plaintext must be less than $p-1~(P=%
\mathbb{Z}
_{p}^{\ast })~$the ElGamal cryptosystem,$~\alpha \in ~\left(
the~chosen~subgroup~of~%
\mathbb{Z}
_{m}^{\ast }\right) $ , the plaintext must be less than $m-1$ $($ $P~=~%
\mathbb{Z}
_{m}/~\{0\})$ in ElGamal Cryptpsystem with Power Fibonacci Sequence module $m
$. In addition, we know that if in ElGamal Cryptpsystem $p$ is a large prime
number , in ElGamal Cryptpsystem with Power Fibonacci Sequence module $m$ , $%
m$ is more large number( $m~=~p_{1}^{e_{1}}.p_{2}^{e_{2}}...p_{k}^{e_{k}}~$%
or $m~=~5p_{1}^{e_{1}}.p_{2}^{e_{2}}...p_{k}^{e_{k}}$ , for each $p_{i}$ $%
\equiv \pm 1({mod}~10)$ is large prime number ). That is, if we choose $%
m~$a composite number by using the power Fibonacci sequence module $m$, we
obtained that for the answer of the question of how this limit in the
ElGamal cryptosystem change in new cryptosystem which we constitute this
limit decrease as $m~$increases. So, ElGamal Cryptpsystem with Power
Fibonacci Sequence module $m~$which we constitute is more advantageous than
ElGamal Cryptpsystem in terms of cryptography.\qquad 

\section{An application of ElGamal Cryptosystem with Power Fibonacci
Sequence Module $M$}

Small examples illustrates following.

\begin{example}
Suppose $m~=~209~$which provides Theorem 1 and so~%
\begin{equation*}
\mathbb{Z}
_{209}^{\ast }=~\{x\in 
\mathbb{Z}
_{209}:(x,209)=1~\}.
\end{equation*}

Module $m~=~209$ , there are four power Fibonacci sequences:

1, 15, 16, 31, 47, 78, 125, 203, 119, 113, 23, 136, 159, 86, 36, 122, 158,
71, 20, 91, 111, 202, 104, 97, 201, 89, 81, 170, 42, 3, 45, 48, 93, 141, 25,
166, 191, 148, 130, 69, 199, 59, 49, 108, 157, 56, 4, 60, 64, 124, 188, 103,
82, 185, 58, 34, 92, 126, 9,135, 144, 70, 5, 75, 80, 155, 26, 181, 207, 179,
177, 147, 115, 53, 168, 12, 180, 192, 163, 146, 100, 37, 137, 174, 102, 67,
169, 27, 196, 14, 1, 15, . . .

1, 81, 82, 163, 36, 199, 26, 16, 42, 58, 100, 158, 49, 207, 47, 45, 92, 137,
20, 157, 177, 125, 93, 9, 102, 111, 4, 115, 119, 25, 144, 169, 104, 64, 168,
23, 191, 5, 196, 201, 188, 180, 159, 130, 80, 1, 81, \ldots

1, 129, 130, 50, 180, 21, 201, 13, 5, 18, 23, 41, 64, 105, 169, 65, 25, 90,
115, 205, 111, 107, 9, 116, 125, 32, 157, 189, 137, 117, 45, 162, 207, 160,
158, 109, 58, 167, 16, 183, 199, 173, 163, 127, 81, 208, 80, 79, 159, 29,
188, 8, 196, 204, 191, 186, 168, 145, 104, 40, 144, 184, 119, 94, 4, 98,
102, 200, 93, 84, 177, 52, 20, 72, 92, 164, 47, 2, 49, 51, 100, 151, 42,
193, 26, 10, 36, 46, 82, 128, 1, 129, \ldots\ and

1, 195, 196, 182, 169, 142, 102, 35, 137, 172, 100, 63, 163, 17, 180, 197,
168, 156, 115, 62, 177, 30, 207, 28, 26, 54, 80, 134, 5, 139, 144, 74, 9,
83, 92, 175, 58, 24, 82, 106, 188, 85, 64, 149, 4, 153, 157, 101, 49, 150,
199, 140, 130, 61, 191, 43, 25, 68, 93, 161, 45, 206, 42, 39, 81, 120, 201,
112, 104, 7, 111, 118, 20, 138, 158, 87, 36, 123, 159, 73, 23, 96, 119, 6,
125, 131, 47, 178, 16, 194, 1, 195, \ldots\ 

We have choosed one of these power Fibonacci sequences.

Curiously,~\{1, 15, 16, 31, 47, 78, 125, 203, 119, 113, 23, 136, 159, 86,
36, 122, 158, 71, 20, 91, 111, 202, 104, 97, 201, 89, 81, 170, 42, 3, 45,
48, 93, 141, 25, 166, 191, 148, 130, 69, 199, 59, 49, 108, 157, 56, 4, 60,
64, 124, 188, 103, 82, 185, 58, 34, 92, 126, 9,135, 144, 70, 5, 75, 80, 155,
26, 181, 207, 179, 177, 147, 115, 53, 168, 12, 180, 192, 163, 146, 100, 37,
137, 174, 102, 67, 169, 27, 196, 14\} is both a subgroup of $%
\mathbb{Z}
_{209}^{\ast }~$and a power Fibonacci sequence modulo~$209$ . $\alpha ~$is a
primitive element of the chosen subgroup of $%
\mathbb{Z}
_{m}^{\ast }$. So, the primitive element $\alpha ~=~15$.

Let $\lambda =78$ , so~$\beta ~=15^{78}({mod}~209)=163$

Now, suppose $~$that Alice wishes to send the message $x=201$ to Bob. Say $%
k=67$ is the random integer she chooses. Then she computes

\begin{multline*}
y_{1} =\alpha ^{^{k}~}({mod}~m)=15^{^{67}~}({mod}~209)=181 \\
y_{2} =x\beta ^{^{k}~}({mod}~m)=201.163^{67~}({mod}~209)=201.125(%
{mod}~209)=45
\end{multline*}

Alice sends $y=~\left( y_{1}~,y_{2}\right) =\left( 181~,45\right) ~$to Bob.
When Bob receives the ciphertext $y=\left( 181~,45\right) $, he computes

\begin{multline*}
x=y_{2}\left( y_{1}~^{\lambda }\right) ^{-1}({mod}~m)=45.\left(
181^{78}\right) ^{-1}({mod}~209)=~45.\left( 125\right) ^{-1}~({mod}%
~209)\\=~45.102({mod}~209)=201
\end{multline*}

which was the plaintext that Alice encrypted.
\end{example}

\begin{example}
Suppose $m~=~1045~$which provides Theorem 1 and so~$%
\mathbb{Z}
_{1045}^{\ast }=~\{x\in 
\mathbb{Z}
_{1045}:(x,1045)=1~\}.$

Module$~m~=~1045$, there are four power Fibonacci sequences:

1, 338, 339, 677, 1016, 648, 619, 222, 841, 18, 859, 877, 691, 523, 169,
692, 861, 508, 324, 832, 111, 943, 9, 952, 961, 868, 784, 607, 346, 953,
254, 162, 416, 578, 994, 527, 476, 1003, 434, 392, 826, 173, 999, 127, 81,
208, 289, 497, 786, 238, 1024, 217, 196, 413, 609, 1022, 586, 563, 104, 667,
771, 393, 119, 512, 631, 98, 729, 827, 511, 293, 804, 52, 856, 908, 719,
582, 256, 838, 49, 887, 936, 778, 669, 402, 26, 428, 454, 882, 291, 128,
419, 547, 966, 468, 389, 857, 201, 13, 214, 227, 441, 668, 64, 732, 796,
483, 234, 717, 951, 623, 529, 107, 636, 743, 334, 32, 366, 398, 764, 117,
881, 998, 834, 787, 576, 318, 894, 167, 16, 183, 199, 382, 581, 963, 499,
417, 916, 288, 159, 447, 606, 8, 614, 622, 191, 813, 1004, 772, 731, 458,
144, 602, 746, 303, 4, 307, 311, 618, 929, 502, 386, 888, 229, 72, 301, 373,
674, 2, 676, 678, 309, 987, 251, 193, 444, 637, 36, 673, 709, 337, 1, 338,
\ldots\ 

1, 433, 434, 867, 256, 78, 334, 412, 746, 113, 859, 972, 786, 713, 454, 122,
576, 698, 229, 927, 111, 1038, 104, 97, 201, 298, 499, 797, 251, 3, 254,
257, 511, 768, 234, 1002, 191, 148, 339, 487, 826, 268, 49, 317, 366, 683,
4, 687, 691, 333, 1024, 312, 291, 603, 894, 452, 301, 753, 9, 762, 771, 488,
214, 702, 916, 573, 444, 1017, 416, 388, 804, 147, 951, 53, 1004, 12, 1016,
1028, 999, 982, 936, 873, 764, 592, 311, 903, 169, 27, 196, 223, 419, 642,
16, 658, 674, 287, 961, 203, 119, 322, 441, 763, 159, 922, 36, 958, 994,
907, 856, 718, 529, 202, 731, 933, 619, 507, 81, 588, 669, 212, 881, 48,
929, 977, 861, 793, 609, 357, 966, 278, 199, 477, 676, 108, 784, 892, 631,
478, 64, 542, 606, 103, 709, 812, 476, 243, 719, 962, 636, 553, 144, 697,
841, 493, 289, 782, 26, 808, 834, 597, 386, 983, 324, 262, 586, 848, 389,
192, 581, 773, 309, 37, 346, 383, 729, 67, 796, 863, 614, 432, 1, 433, \ldots

1, 613, 614, 182, 796, 978, 729, 662, 346, 1008, 309, 272, 581, 853, 389,
197, 586, 783, 324, 62, 386, 448, 834, 237, 26, 263, 289, 552, 841, 348,
144, 492, 636, 83, 719, 802, 476, 233, 709, 942, 606, 503, 64, 567, 631,
153, 784, 937, 676, 568, 199, 767, 966, 688, 609, 252, 861, 68, 929, 997,
881, 833, 669, 457, 81, 538, 619, 112, 731, 843, 529, 327, 856, 138, 994,
87, 36, 123, 159, 282, 441, 723, 119, 842, 961, 758, 674, 387, 16, 403, 419,
822, 196, 1018, 169, 142, 311, 453, 764, 172, 936, 63, 999, 17, 1016, 1033,
1004, 992, 951, 898, 804, 657, 416, 28, 444, 472, 916, 343, 214, 557, 771,
283, 9, 292, 301, 593, 894, 442, 291, 733, 1024, 712, 691, 358, 4, 362, 366,
728, 49, 777, 826, 558, 339, 897, 191, 43, 234, 277, 511, 788, 254, 1042,
251, 248, 499, 747, 201, 948, 104, 7, 111, 118, 229, 347, 576, 923, 454,
332, 786, 73, 859, 932, 746, 633, 334, 967, 256, 178, 434, 612, 1, 613,
\ldots\ and

1, 708, 709, 372, 36, 408, 444, 852, 251, 58, 309, 367, 676, 1043, 674, 672,
301, 973, 229, 157, 386, 543, 929, 427, 311, 738, 4, 742, 746, 443, 144,
587, 731, 273, 1004, 232, 191, 423, 614, 1037, 606, 598, 159, 757, 916, 628,
499, 82, 581, 663, 199, 862, 16, 878, 894, 727, 576, 258, 834, 47, 881, 928,
764, 647, 366, 1013, 334, 302, 636, 938, 529, 422, 951, 328, 234, 562, 796,
313, 64, 377, 441, 818, 214, 1032, 201, 188, 389, 577, 966, 498, 419, 917,
291, 163, 454, 617, 26, 643, 669, 267, 936, 158, 49, 207, 256, 463, 719,
137, 856, 993, 804, 752, 511, 218, 729, 947, 631, 533, 119, 652, 771, 378,
104, 482, 586, 23, 609, 632, 196, 828, 1024, 807, 786, 548, 289, 837, 81,
918, 999, 872, 826, 653, 434, 42, 476, 518, 994, 467, 416, 883, 254, 92,
346, 438, 784, 177, 961, 93, 9, 102, 111, 213, 324, 537, 861, 353, 169, 522,
691, 168, 859, 1027, 841, 823, 619, 397, 1016, 368, 339, 707, 1, 708,
\ldots\ 

We have choosed one of these power Fibonacci sequences.

Curiously, \{1, 338, 339, 677, 1016, 648, 619, 222, 841, 18, 859, 877, 691,
523, 169, 692, 861, 508, 324, 832, 111, 943, 9, 952, 961, 868, 784, 607,
346, 953, 254, 162, 416, 578, 994, 527, 476, 1003, 434, 392, 826, 173, 999,
127, 81, 208, 289, 497, 786, 238, 1024, 217, 196, 413, 609, 1022, 586, 563,
104, 667, 771, 393, 119, 512, 631, 98, 729, 827, 511, 293, 804, 52, 856,
908, 719, 582, 256, 838, 49, 887, 936, 778, 669, 402, 26, 428, 454, 882,
291, 128, 419, 547, 966, 468, 389, 857, 201, 13, 214, 227, 441, 668, 64,
732, 796, 483, 234, 717, 951, 623, 529, 107, 636, 743, 334, 32, 366, 398,
764, 117, 881, 998, 834, 787, 576, 318, 894, 167, 16, 183, 199, 382, 581,
963, 499, 417, 916, 288, 159, 447, 606, 8, 614, 622, 191, 813, 1004, 772,
731, 458, 144, 602, 746, 303, 4, 307, 311, 618, 929, 502, 386, 888, 229, 72,
301, 373, 674, 2, 676, 678, 309, 987, 251, 193, 444, 637, 36, 673, 709,
337\} is both a subgroup of $%
\mathbb{Z}
_{1045}^{\ast }~$and a power Fibonacci sequence modulo~$1045$ . $\alpha ~$is
a primitive element of the chosen subgroup of $%
\mathbb{Z}
_{m}^{\ast }$. So, the primitive element $\alpha ~=~338$.

Let $\lambda =547$ , so~$\beta ~=338^{547}({mod}~1045)=222.$

Now, suppose $~$that Alice wishes to send the message $x=1001$ to Bob. Say $%
k=162$ is the random integer she chooses. Then she computes

\begin{multline*}
y_{1} =\alpha ^{^{k}~}({mod}~m)=338^{^{162}~}({mod}~1045)=229 \\
y_{2}=x\beta ^{^{k}~}({mod}~m)=1001.222^{162~}({mod}%
~1045)=1001.609({mod}~1045)=374
\end{multline*}

Alice sends $y=~\left( y_{1}~,y_{2}\right) =\left( 229~,374\right) ~$to Bob.
When Bob receives the ciphertext $y=\left( 229~,374\right) $, he computes
\begin{multline*}
x=y_{2}\left( y_{1}~^{\lambda }\right) ^{-1}({mod}~m)=374.\left(
229^{547}\right) ^{-1}({mod}~1045)=~374.\left( 609\right) ^{-1}~({%
	mod}~1045)\\=~374.894({mod}~209)=1001
\end{multline*}

which was the plaintext that Alice encrypted.
\end{example}

\section{Conclusion}

In this study, we have adapted to public key cryptography power Fibonacci
sequence module $m$. To do this, we have examined power Fibonacci sequence
module $m~$and we have obtained that a power Fibonaci sequence modulo $m~$%
constitutes a cyclic and multiplicative group whose order is a divisor of $%
\varphi (m)~$at the same time. So, we have obtained that we are able to
reconstitute Discrete Logarithm problem by using the power Fibonacci
sequence module $m$ . Thus, we got new difficult problem, and we called this
problem as Discrete Logarithm problem with Power Fibonacci Sequence module $%
m $ , then we have constructed a new cryptosystem based on the new problem
similar to ElGamal Crptosystem, and we called the new cryptographic system
as ElGamal Cryptpsystem with Power Fibonacci Sequence module $m$ .

In addition one of two limits in the ElGamal cryptosystem is that the
plaintext must be less than $p-1$\cite{Hwang}. We have compared that ElGamal
Cryptpsystem and ElGamal Cryptpsystem with Power Fibonacci Sequence module $%
m $ in terms of this limit and we have obtained that while $\alpha \in 
\mathbb{Z}
_{p}^{\ast }$, the plaintext must be less than $p-1(P~=%
\mathbb{Z}
_{p}^{\ast })$ in the ElGamal cryptosystem, $\alpha \in ~\left(
the~chosen~subgroup~of~%
\mathbb{Z}
_{m}^{\ast }\right) $ , the plaintext must be less than $m-1$ $(P~=~%
\mathbb{Z}
_{m}/~\{0\})$ in ElGamal Cryptpsystem with Power Fibonacci Sequence module $%
m $. Moreover, we know that if in ElGamal Cryptpsystem $p$ is a large prime
number , in ElGamal Cryptpsystem with Power Fibonacci Sequence module $m$ , $%
m$ is more large number ($m~=~p_{1}^{e_{1}}.p_{2}^{e_{2}}...p_{k}^{e_{k}}~$%
or $m~=~5p_{1}^{e_{1}}.p_{2}^{e_{2}}...p_{k}^{e_{k}}$ , for each $p_{i}$ $%
\equiv \pm 1({mod}~10)$ is large prime number . That is, if we choose $%
m~$a composite number by using the power Fibonacci sequence module $m$ , we
obtained that this limit decrease as $m~$increases.

So, by means of the new cryptosystem, we have made that the cryptosystem
which is used only in prime modulus is also usable for composite modulus and
also the new cryptosystem which we defined is more advantages than ElGamal
Cryptpsystem in terms of crptography. Because, in comparison with ElGamal
Cryptpsystem, for one who doesn't know the private key the number of data
which must try to understand the message increase in ElGamal Cryptpsystem
with Power Fibonacci Sequence module $m$.

\end{document}